\documentclass{appolb}
\usepackage{graphicx}
\RequirePackage{lineno}

\def\beq{\begin{equation}}
\def\eeq{\end{equation}}
\def\bea{\begin{eqnarray}}
\def\eea{\end{eqnarray}}

\newcommand{\Lb}{\left(}
\newcommand{\Rb}{\right)}
\newcommand \sqn{$\sqrt{s_{_{NN}}}$ } 

\usepackage[rightcaption]{sidecap}
\usepackage{graphicx,color,amsmath,amssymb}

\usepackage{amssymb}

\usepackage[figuresright]{rotating}

\begin{document}

\title{PHENIX measurements of low momentum direct photon radiation from large and small systems in (ultra)relativistic heavy ion collisions: direct photon scaling\thanks{Presented at the XIIIth Workshop on Particle Correlations and Femtoscopy \\ (WPCF 2018), Krakow, Poland, May 22-26, 2018.}%
}

\author{Vladimir Khachatryan for the PHENIX Collaboration %
\address{Department of Physics and Astronomy, Stony Brook University, Stony Brook, New York 11794-3800, USA}
}

\maketitle
\begin{abstract}
The PHENIX collaboration has measured low momentum direct photon radiation in Au+Au collisions at 200\,GeV, 62.4\,GeV and 39\,GeV, in Cu+Cu at 200\,GeV as well as in p+p, p+Au and d+Au at $\sqrt{s_{NN}} = $200\,GeV. In these measurements PHENIX has discovered a large excess over the scaled p+p yield of direct photons in A+A collisions, and a non-zero excess, observed within systematic uncertainties, over the scaled p+p yield in central p+A collisions. Another finding is that at low-$p_{T}$ the integrated yield of direct photons, $dN_{\gamma}/dy$, from large systems shows a behavior of universal scaling as a function of the charged-particle multiplicity, $(dN_{ch}/d\eta)^{\alpha}$, with $\alpha = 1.25$, which means that the photon production yield increases faster than the charged-particle multiplicity.
\end{abstract}



\setlength{\linenumbersep}{6pt}

\section{Introduction}
\label{intro}
Direct photons determine the excess yield, which one obtains by subtracting the hadronic decay photon yield (mostly from $\pi^{0}$ and $\eta$ decays) from the total observed photon yield. By measuring these photons, we can study the strongly interacting medium produced in (ultra)relativistic heavy ion collisions, and gain information on the properties and dynamics of the produced matter integrated over space and time. The direct photons possibly originate from the hot fireball of the Quark-Gluon Plasma (QGP), late hadronic phase as well as from initial hard scattering processes like QCD Compton scattering among the incoming and outgoing partons.
 
After PHENIX measured large invariant yield and large anisotropy of low momentum direct photons in Au+Au collisions at \sqn = 200\,GeV \cite{Adare:2008qk,Bannier:2014,Bannier:2016}, a challenging problem arose, commonly referred to as ``thermal photon puzzle", where different theoretical models encounter difficulties when they are used to describe these two quantities simultaneously (though there is also some progress in recent years \cite{vanHees:2014ida,Paquet:2015lta,Kim:2016ylr} on this matter). For resolving this puzzle, PHENIX continues measuring low momentum direct photons in large and small collision systems. These past and new measurements revealed very interesting findings, which we report in this proceedings.

\section{Some results on direct photon elliptic flow from large systems and on direct photon $p_{T}$ spectra from large/small systems}
\label{results}
\begin{figure}[h!]
\vspace{-3.5mm}
\begin{center}
\hspace{0.0\textwidth}
\vspace{0.0\textwidth}
   {\includegraphics[width=0.485\textwidth]{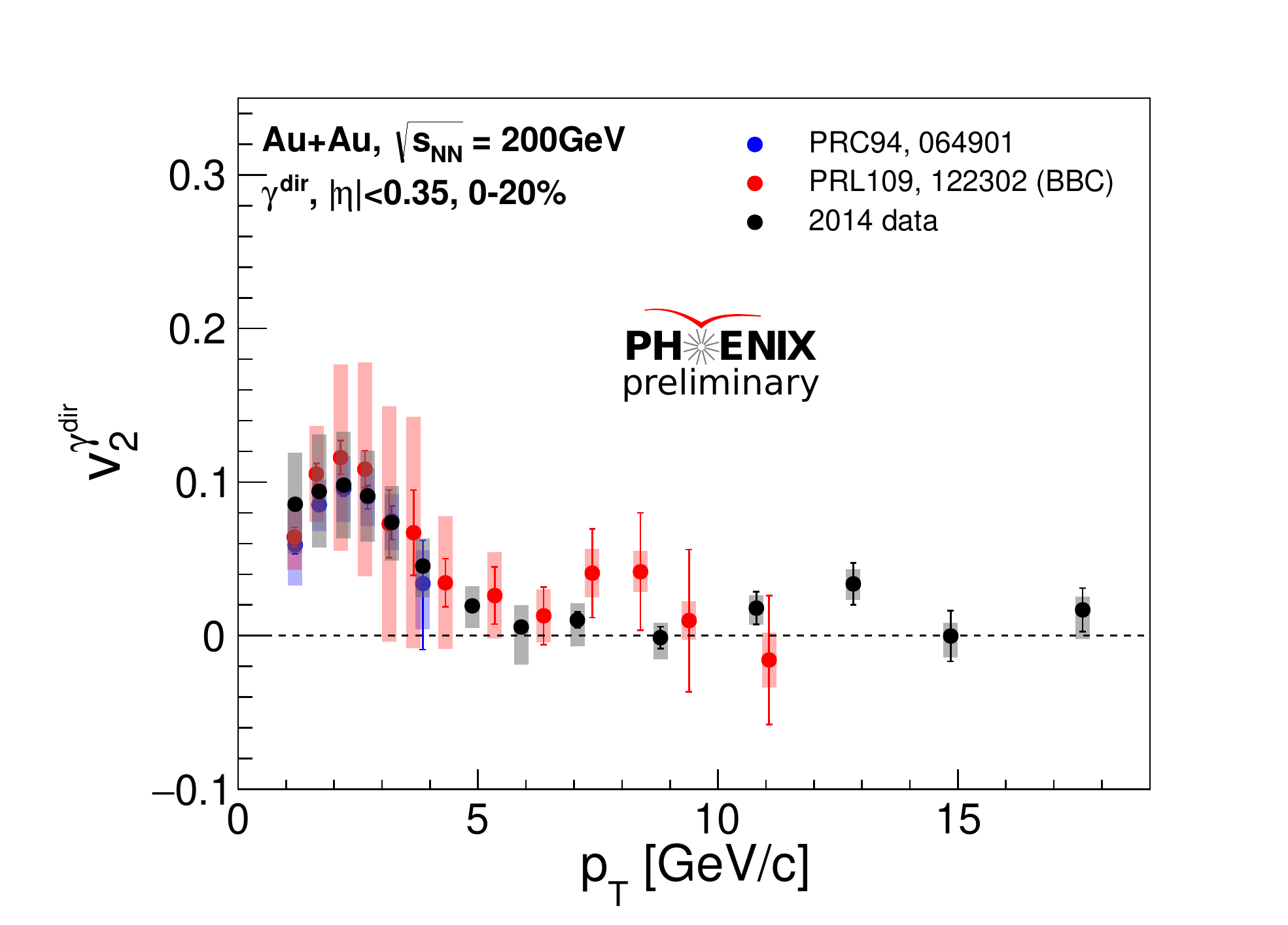} \label{fig:Fig1b}}
\hspace{-0.05\textwidth}
\vspace{0.0\textwidth}
   {\includegraphics[width=0.485\textwidth]{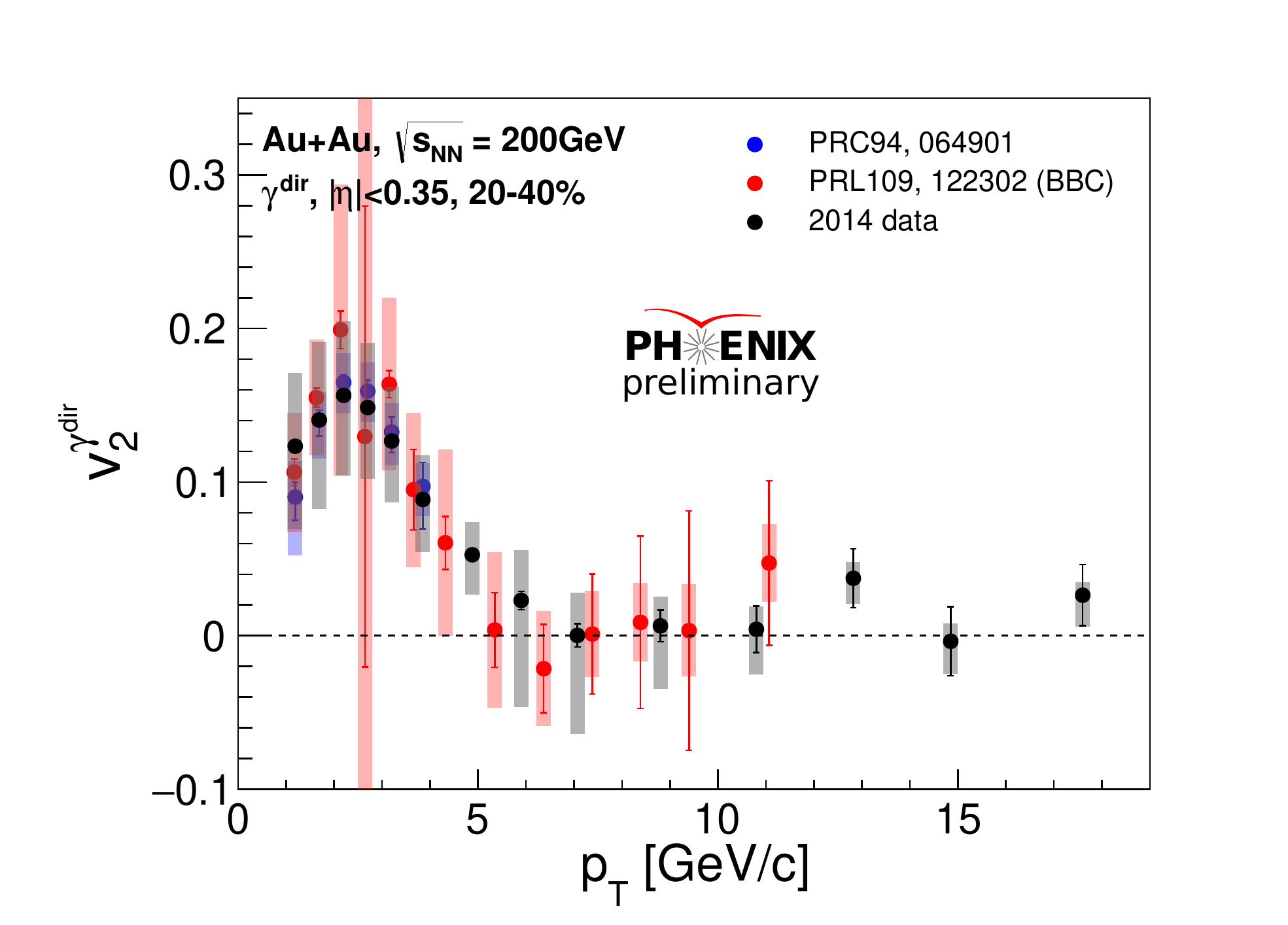} \label{fig:Fig1c}}
\end{center}
\vspace{-6.0mm}
\caption{(Left) Direct photon $v_{2}$ vs. $p_{T}$ in 0-20\% compared with the published results from \cite{Bannier:2016}. (Right) Direct photon $v_{2}$ vs. $p_{T}$ in 20-40\% compared with the published results from \cite{Bannier:2016}. The new results are based upon external conversions with VTX.}
\label{fig:fig_v2}
\end{figure}

\begin{figure}[h!]
\vspace{0.0mm}
\begin{center}
\hspace{0.0\textwidth}
\vspace{0.0\textwidth}
   {\includegraphics[width=0.365\textwidth]{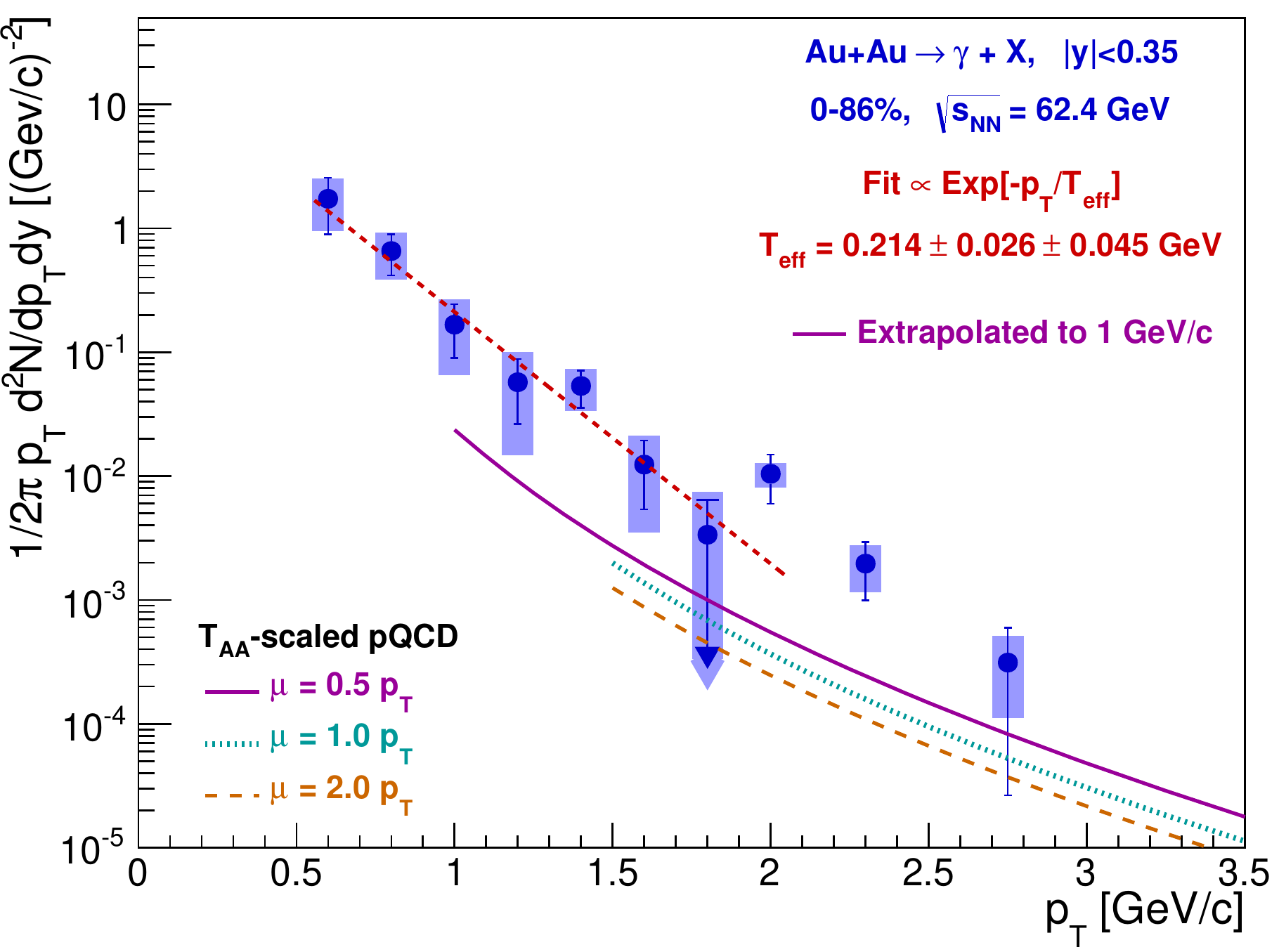} \label{fig:Fig1b}}
\hspace{-0.03\textwidth}
\vspace{0.0\textwidth}
   {\includegraphics[width=0.25\textwidth, height=0.22\textheight]{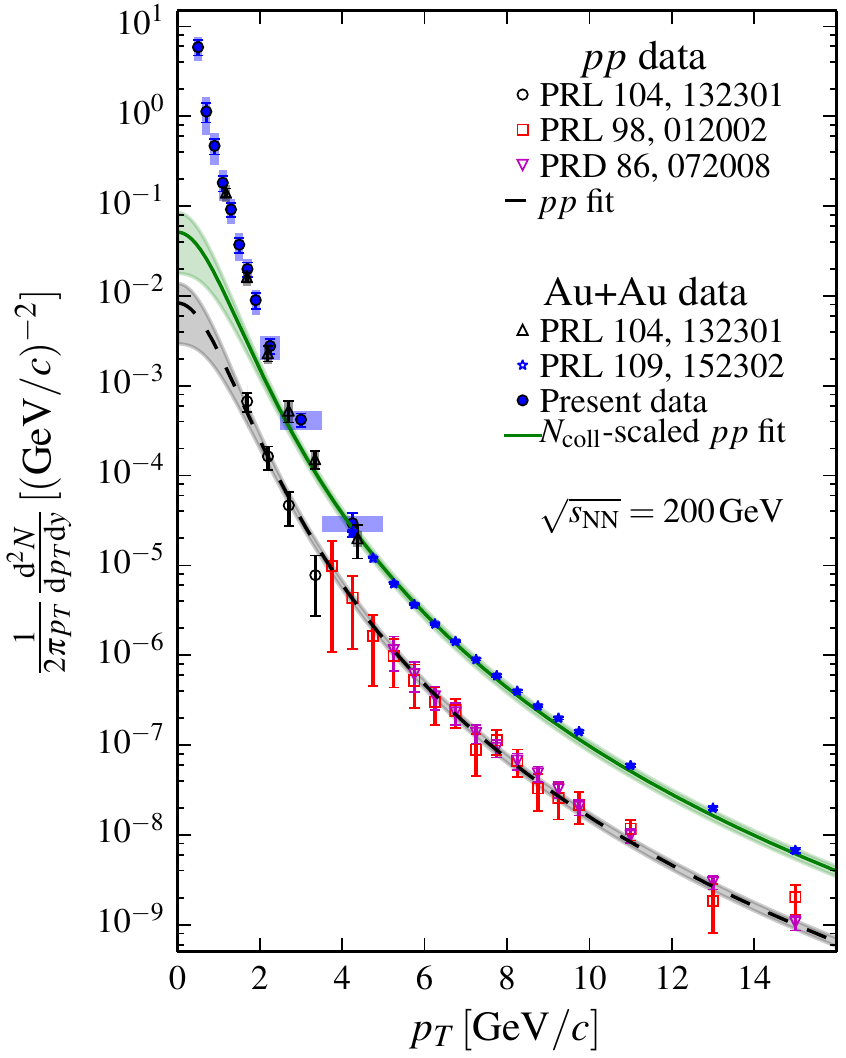} \label{fig:Fig1a}}
\hspace{-0.02\textwidth}
\vspace{0.0\textwidth}
   {\includegraphics[width=0.365\textwidth]{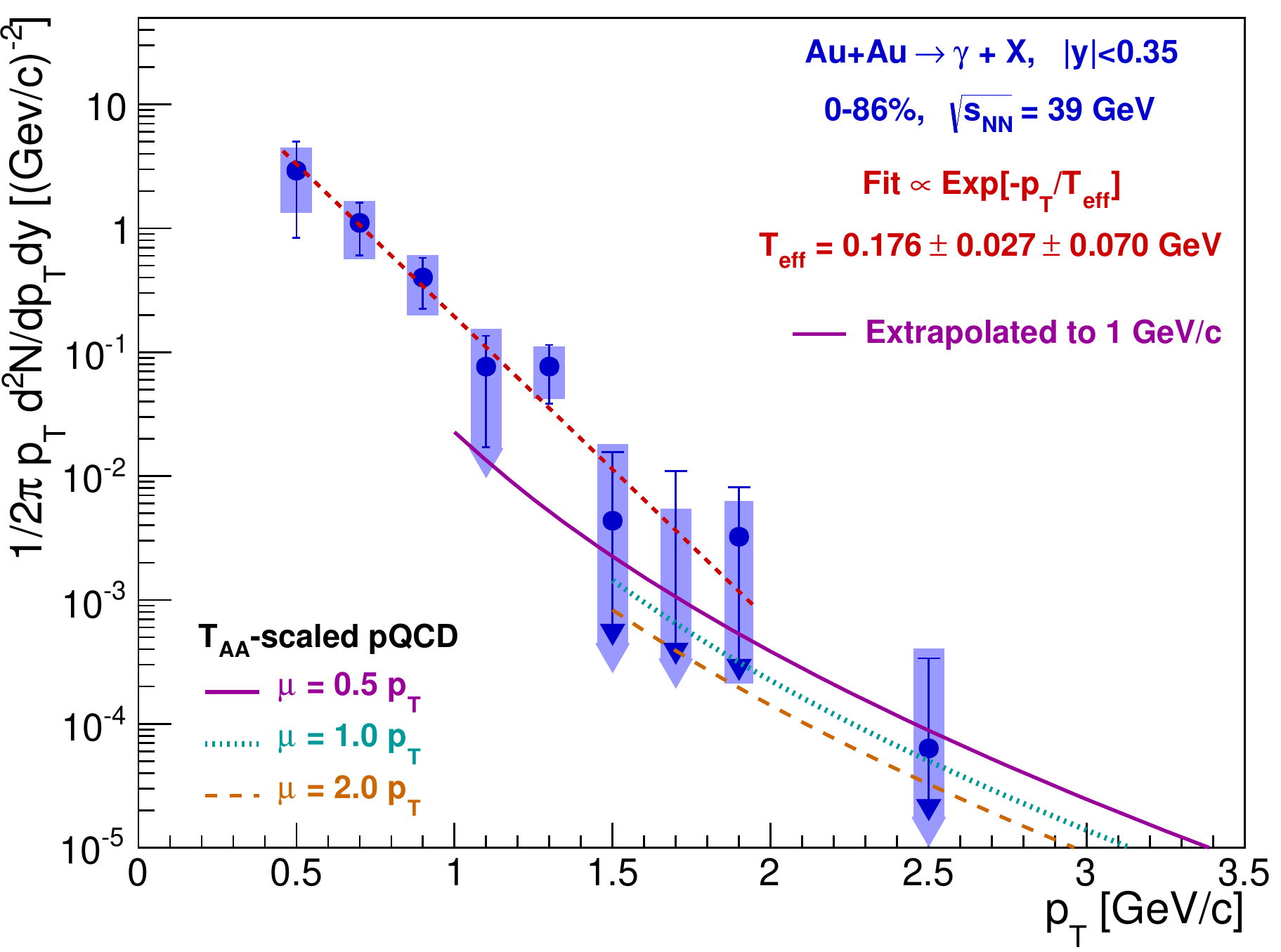} \label{fig:Fig1c}}
\end{center}
\vspace{-10.0mm}
\caption{Direct photon $p_{T}$ spectra in Au+Au at \sqn = 200\,GeV \cite{Bannier:2014} (central panel), and in Au+Au at \sqn = 62.4\,GeV and 39\,GeV \cite{Adare:2018} (left and right panels). All data are in minimum bias and from external conversions with HBD.}
\label{fig:fig_large}
\end{figure}

In this section we represent some of the PHENIX recent and previously published results on direct photon measurements. Fig.\,\ref{fig:fig_v2} shows two plots on direct photon elliptic flow ($v_{2}$) in two centrality bins, from an ongoing analysis in Au+Au at \sqn = 200 GeV obtained based upon an external conversion method. Fig.\,\ref{fig:fig_large} shows $p_{T}$ spectra, obtained with another external conversion method, for minimum bias data samples in Au+Au at \sqn = 62.4\,GeV and 39\,GeV \cite{Adare:2018}, including also the previously published result in Au+Au at 200 GeV \cite{Bannier:2014}. In the external conversion method the photons are measured through their conversions to $e^{+}e^{-}$ pairs at the VTX or HBD in the PHENIX detector system, and the fraction of direct photons is determined after tagging photons from neutral pion decays. Comparing these data to $N_{coll}$ scaled p+p fit (or pQCD calculations) one finds a significant excess over the scaled p+p yield of low-$p_{T}$ direct photons in all three systems. 

\begin{figure}[h!]
\vspace{0.0mm}
\begin{center}
\hspace{0.0\textwidth}
\vspace{0.0\textwidth}
   {\includegraphics[width=0.3\textwidth]{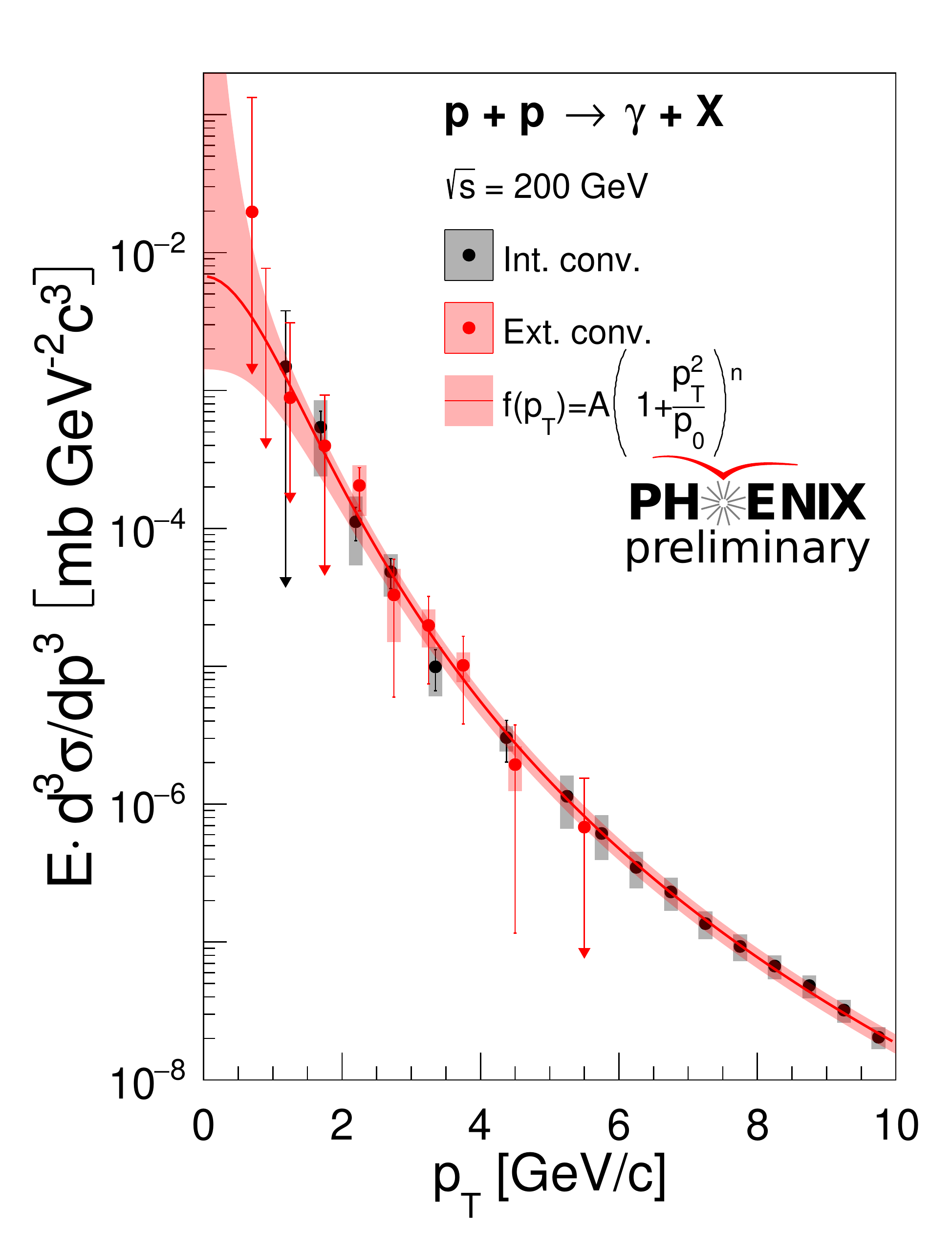} \label{fig:Fig2a}}
\hspace{0.0\textwidth}
\vspace{0.0\textwidth}
   {\includegraphics[width=0.31\textwidth, height=0.27\textheight]{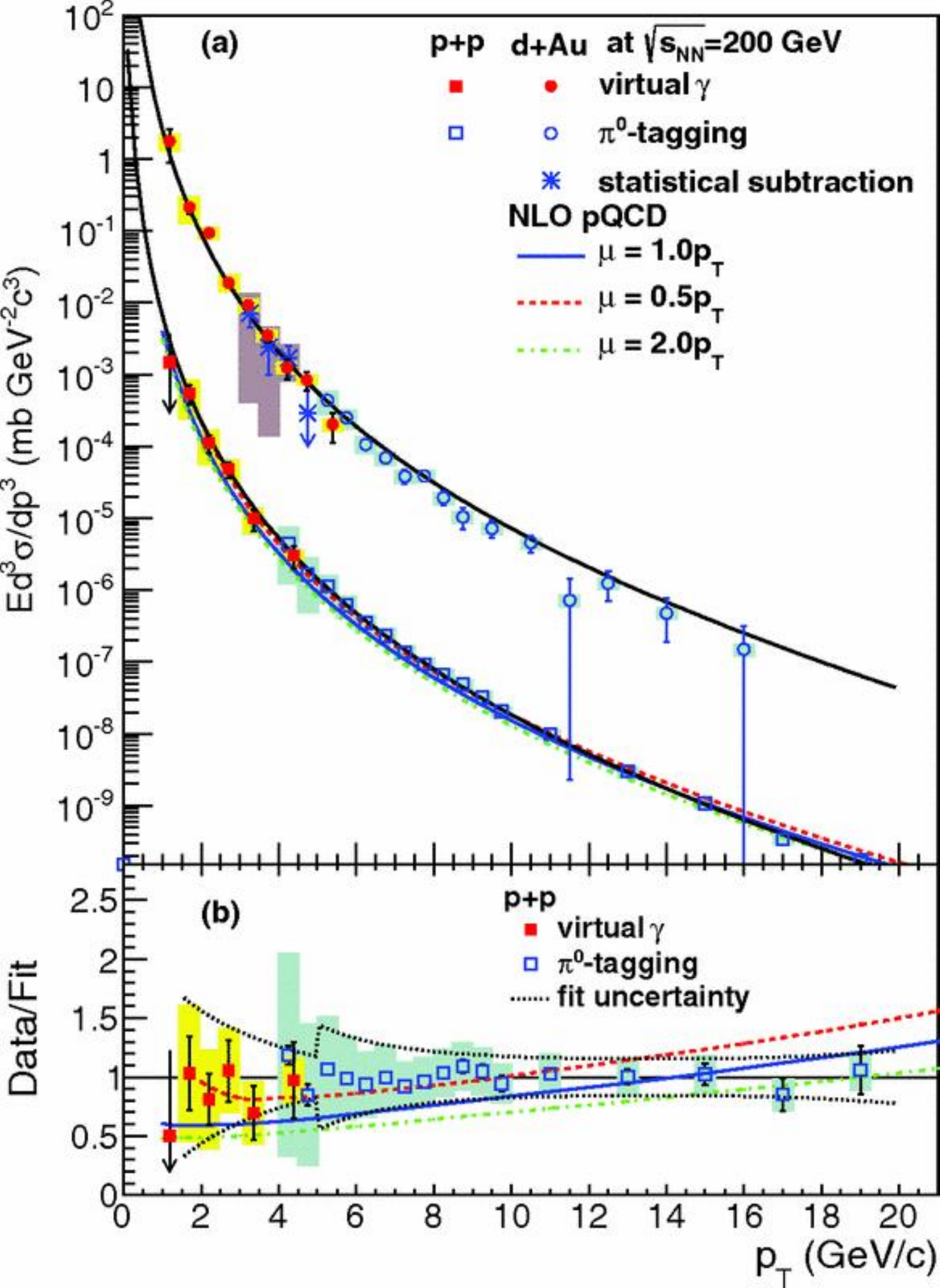} \label{fig:Fig2b}}
\hspace{0.0\textwidth}
\vspace{0.0\textwidth}
   {\includegraphics[width=0.3\textwidth]{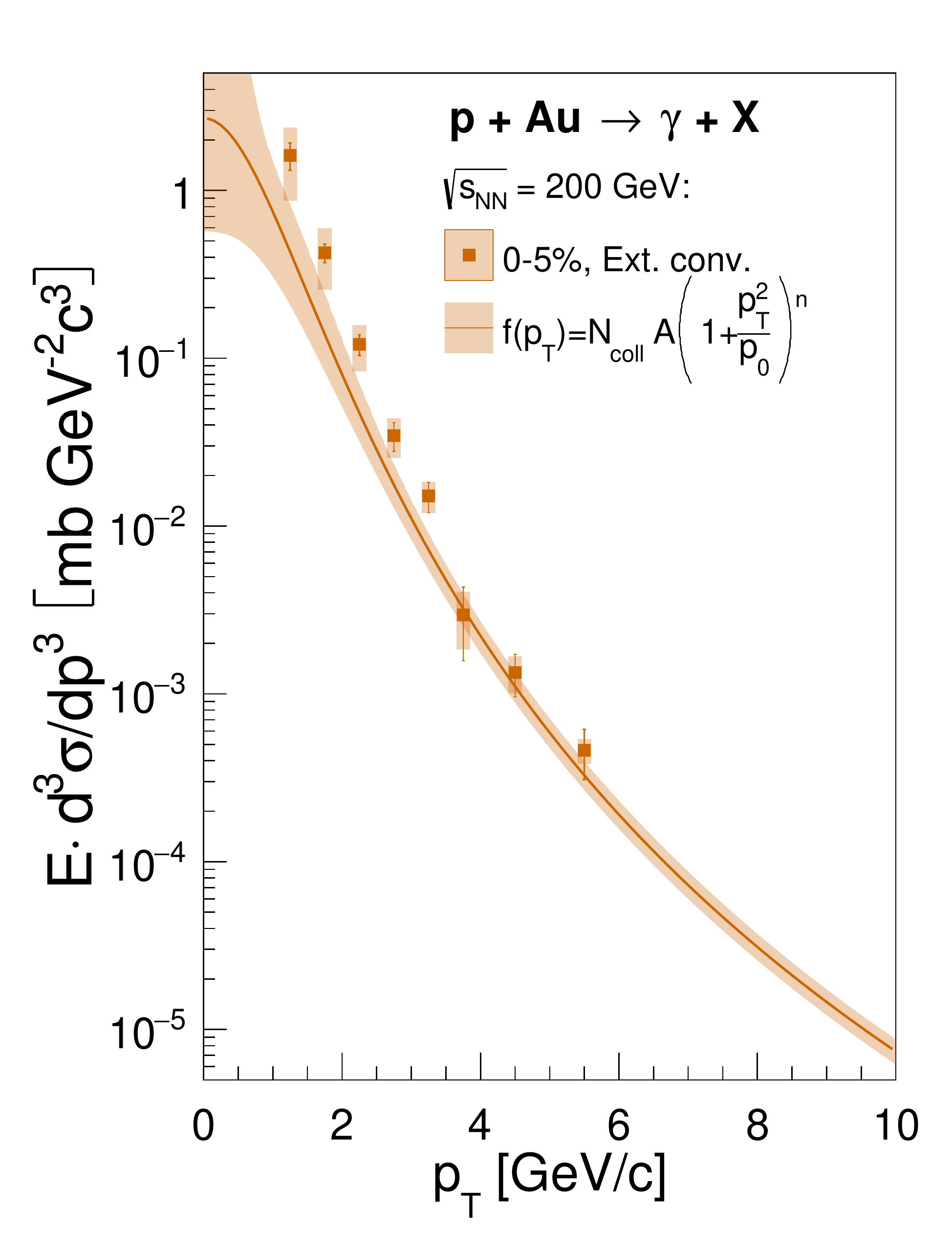} \label{fig:Fig2c}}
\end{center}
\vspace{-5.0mm}
\caption{The left and right panels show direct photon $p_{T}$ spectra in p+p and in the central 0-5\% p+Au collisions at 200\,GeV (external conversions with VTX). The PHENIX previously measured p+p and minimum bias d+Au cross sections from an internal conversion method \cite{Adare:2013} are included in this figure and shown in the left panel (with p+p), and in the central panel (with p+p and d+Au).}
\label{fig:fig_small}
\end{figure}

With the external conversion method PHENIX recently also measured low momentum direct photons in p+p and p+A collisions (shown in Fig.\,\ref{fig:fig_small}). Within systematic uncertainties, the observed a non-zero excess yield ($\sim$ one sigma) in central p+Au collisions above the scaled p+p fit may come from the possible production of QGP droplets in small central systems.

\section{Scaling properties of direct photons}
\label{scaling}
\begin{SCfigure}[][h!]
\vspace{0.0mm}
   \includegraphics[width=0.425\textwidth]{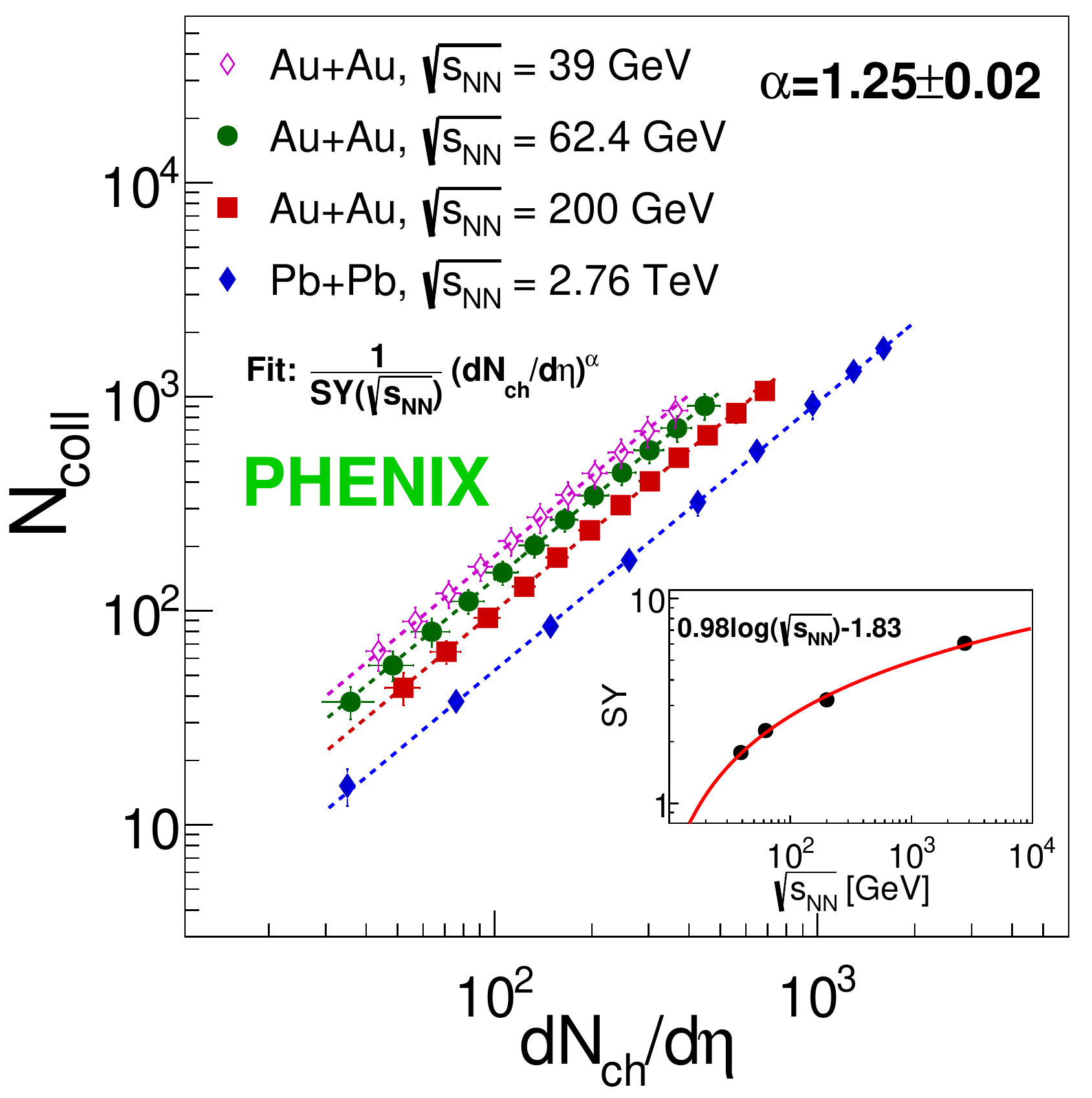}
\hspace{0.0mm}
\vspace{-0.5mm}
\caption{$N_{coll}$ vs. $dN_{ch}/d\eta|_{\eta=0}$ for given four beam energies. The small box on the bottom right shows data demonstrating a scaling between $N_{coll}$ and $dN_{ch}/d\eta$ of the form:
$
N_{coll} = \frac{1}{SY\!(\sqrt{s_{_{NN}}})} \Lb \frac{dN_{ch}}{d\eta} \Rb^{\alpha},
$
where the specific yield is a function logarithmically increasing with \sqn:
$
\mbox{SY}\!(\sqrt{s_{_{NN}}}) = 0.98\!\cdot\!\log(\sqrt{s_{_{NN}}}) - 1.83.
$
{\color{white} ................................................................................
}}
\label{fig:fig_scaling1}
\end{SCfigure}

For a given center-of-mass energy one can compare data from different centrality classes (or system size) using number of participants, $N_{part}$, or the number of binary collisions, $N_{coll}$. But this way is not useful when we compare data at different energies. Instead, we use charged-particle  multiplicity, $dN_{ch}/d\eta$, which itself has an interesting scaling behavior with $N_{coll}$ shown in Fig.\,\ref{fig:fig_scaling1}. Here $N_{coll}$ scales like $(dN_{ch}/d\eta)^{\alpha}$ for all center-of-mass energies with a logarithmically slowly increasing function called specific yield, SY. Four datasets are simultaneously fitted by a power-law, with vertical and horizontal uncertainties of $N_{coll}$ and  $dN_{ch}/d\eta|_{\eta=0}$, respectively \cite{Adare:nch,Aamodt:2011nch}. Then we get $\alpha =1.25 \pm 0.02$. For other details see the caption of Fig.\,\ref{fig:fig_scaling1}.

\vspace*{0.0mm}
\begin{figure}[h!]
\vspace{-1.0mm}
\begin{center}
   {\includegraphics[width=0.85\textwidth]{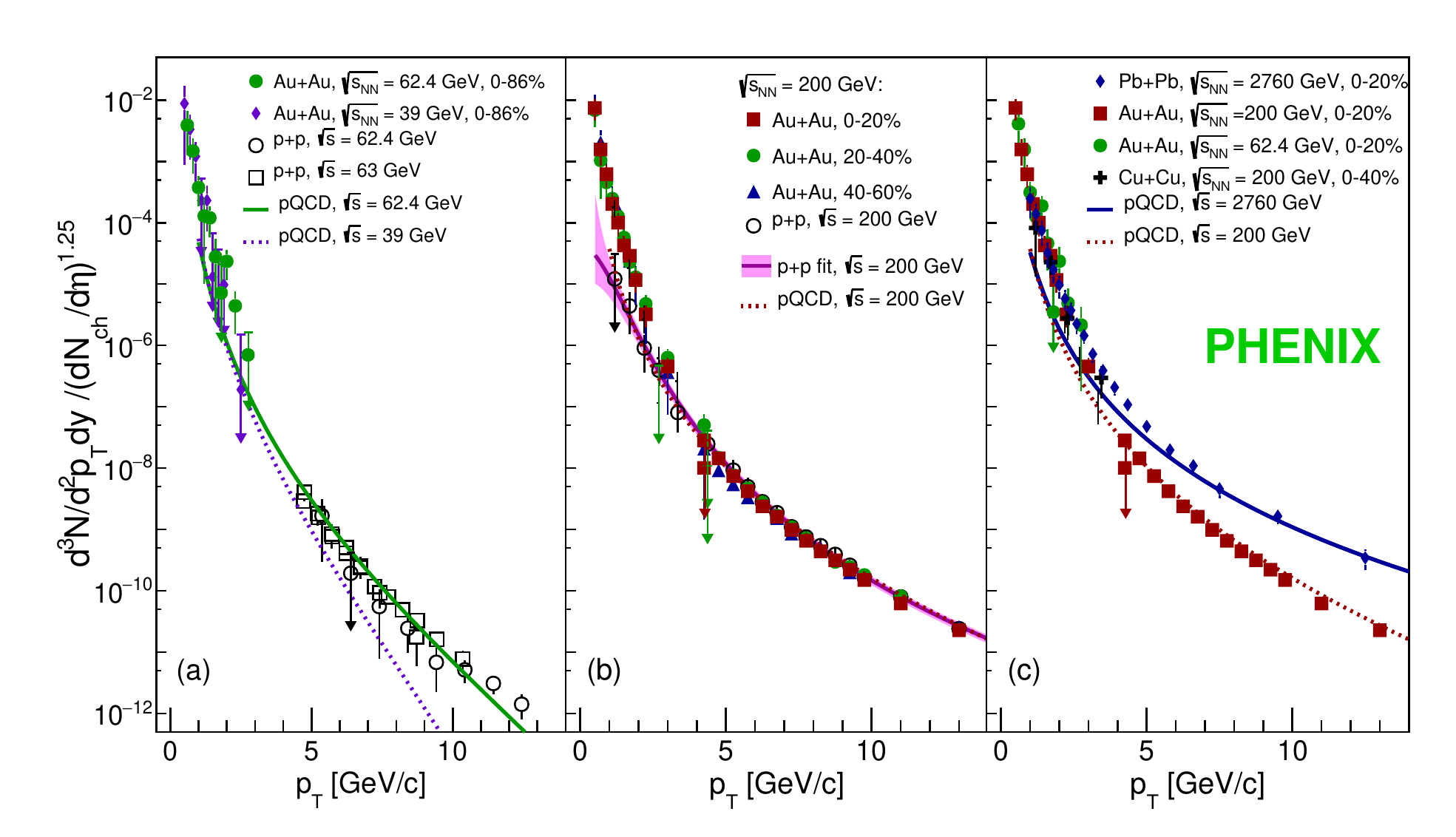}}
\end{center}
\vspace{-7.5mm}
\caption{This three-panel plot is from \cite{Adare:2018} showing the direct photon spectra normalized by $(dN_{ch}/d\eta)^{1.25}$. The comparison is shown for Au+Au data in minimum bias collisions at 62.4 GeV and 39 GeV in the panel (a); for Au+Au data in three centrality bins at 200 GeV in the panel (b); and for different A+A systems at four beam energies in the panel (c). The panels (a) and (b) also show p+p data, and all the panels show perturbative QCD calculations at respective energies.}
\label{fig:fig_scaling2}
\end{figure}

Thereby, one can scale the direct photon yield by $(dN_{ch}/d\eta)^{\alpha}$, which for a specific \sqn is equivalent to $N_{coll}$. Let us take, e.g., the photon spectra in minimum bias Au+Au collisions at 62.4 and 39\,GeV with pQCD curves from Fig.\,\ref{fig:fig_large}, and normalize them by $(dN_{ch}/d\eta)^{\alpha}$. It results in the data falling on top of each other at low-$p_{T}$ as shown in the panel (a) of Fig.\,\ref{fig:fig_scaling2}. As expected at high-$p_{T}$ the p+p data coincide with the pQCD calculations within the quoted uncertainties. In the panel (b) all Au+Au data at 200\,GeV are on top of each other at high- and low-$p_{T}$, and at low-$p_{T}$ they are distinctly above the p+p data, fit and pQCD. In (c) the data are compared for different \sqn from 62.4\,GeV to 2760\,GeV. Again all the data coincide at low-$p_{T}$, while at high-$p_{T}$ we see the expected difference with \sqn and $N_{coll}$ scaling. 

In  Fig.\,\ref{fig:fig_scaling2} all error bars are the quadratic sum of the systematic and statistical uncertainties. Uncertainties on $dN_{ch}/d\eta$ are not included. All normalized data, p+p fit, pQCD curves are from \cite{Adare:2018}, and they are obtained with Au+Au data from \cite{Adare:2008qk,Bannier:2014,Afanasiev:2012}, Pb+Pb data from \cite{Adam:2016}, p+p data at 200\,GeV from \cite{Adare:2013}, at 62.4\,GeV from \cite{Angelis:1980}, at 63\,GeV \cite{Angelis:1989,Akesson:1990}, the empirical fit to the p+p data at 200\,GeV from \cite{Adare:2018}, the pQCD calculations at different beam energies from \cite{Paquet:2015lta,Paquet:2015}, and the data on $dN_{ch}/d\eta$ from \cite{Adare:nch,Aamodt:2011nch}.

\begin{figure}[h!]
\vspace{-2.5mm}
\begin{center}
\hspace{0.0\textwidth}
\vspace{0.0\textwidth}
   {\includegraphics[width=0.475\textwidth]{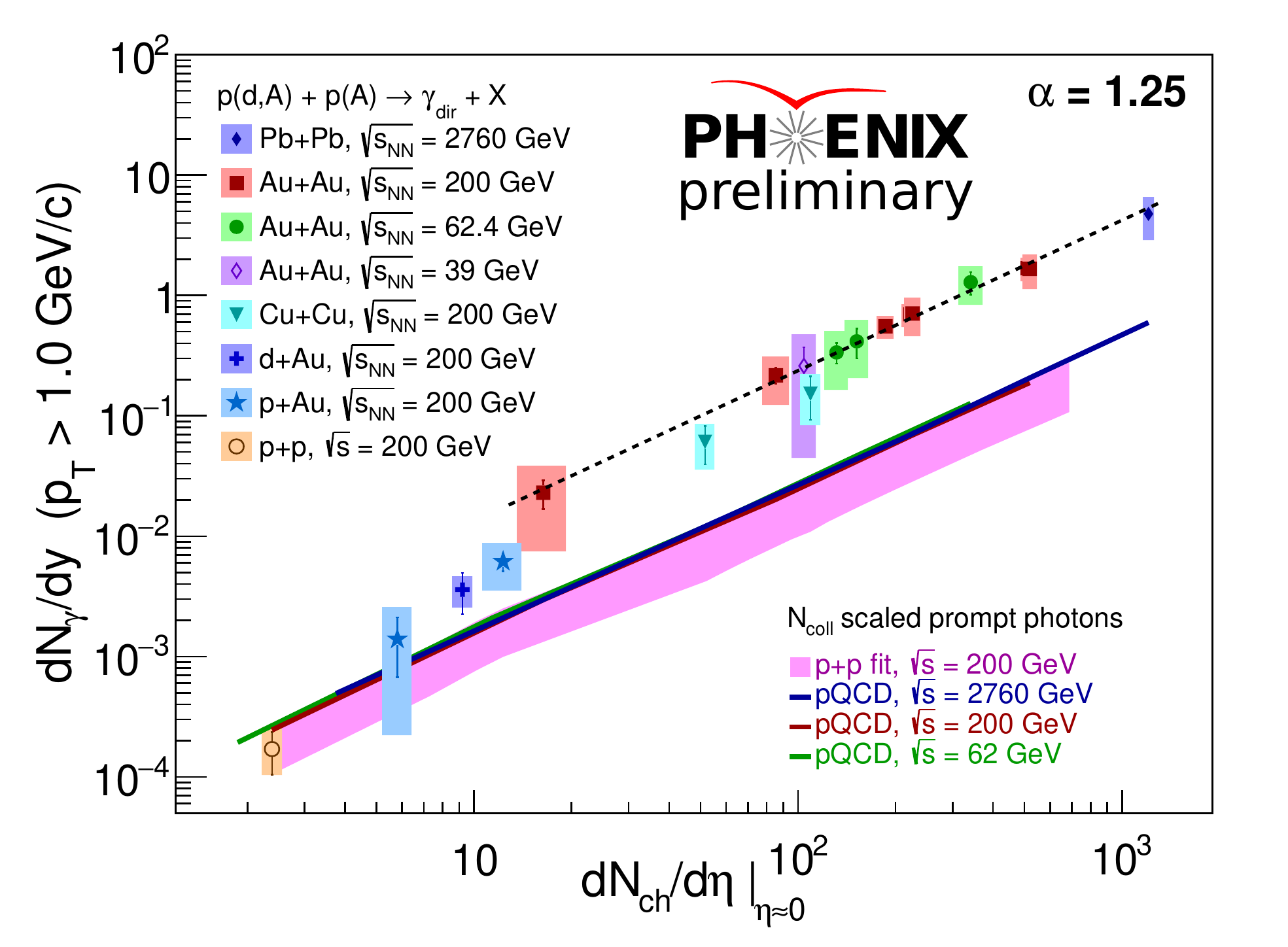} \label{fig:Fig5a}}
\hspace{0.00\textwidth}
\vspace{0.0\textwidth}
   {\includegraphics[width=0.475\textwidth]{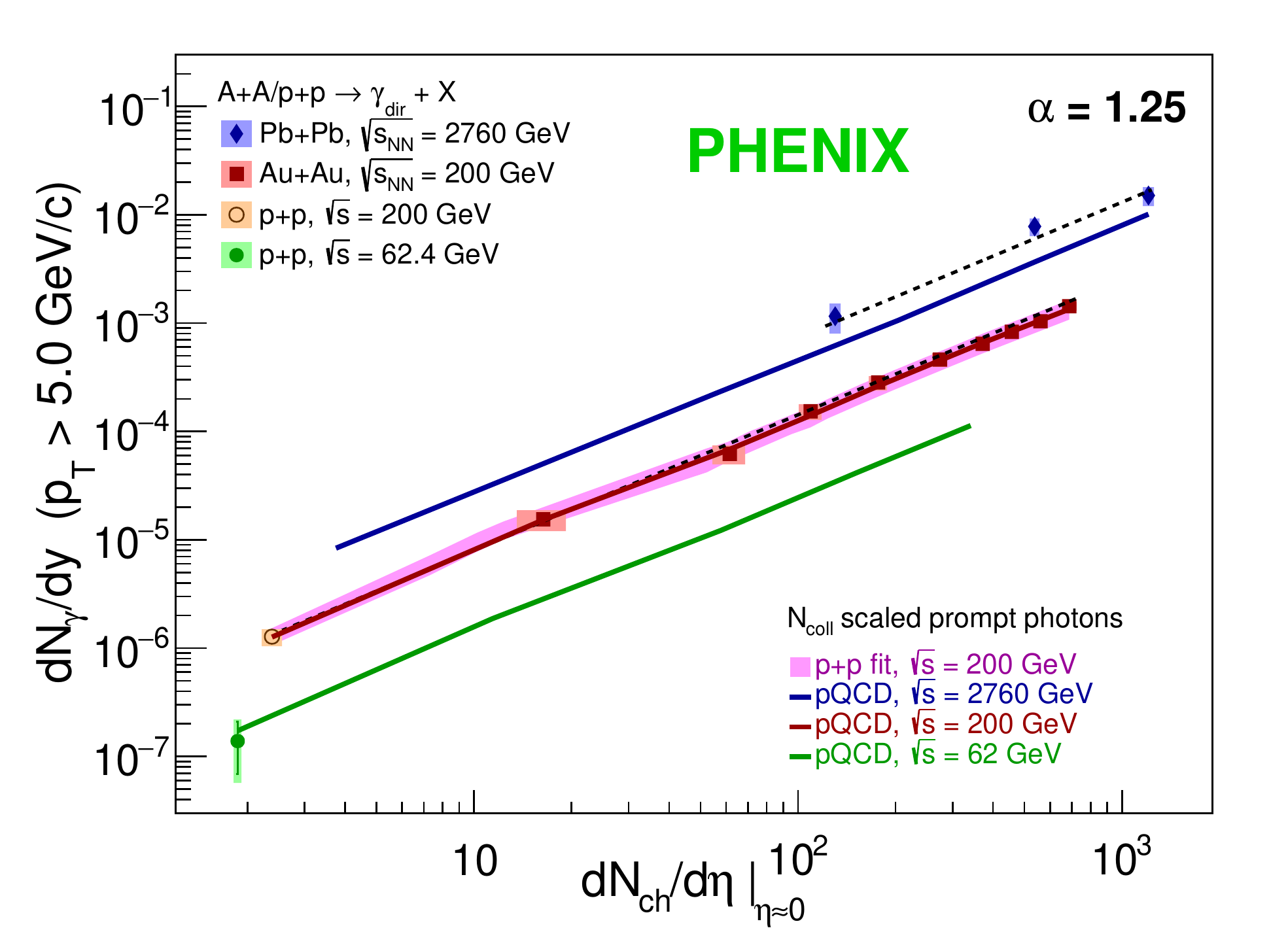} \label{fig:Fig5b}}
\end{center}
\vspace{-6.0mm}
\caption{
The left plot shows the direct photon yield, integrated $>1.0$\,GeV/c in $p_{T}$, vs. $dN_{ch}/d\eta$, for five A+A datasets at different collision energies and for one p+Au, one d+Au and one p+p datasets at 200\,GeV (some of the unintegrated data are shown in Fig.\,\ref{fig:fig_large} and Fig.\,\ref{fig:fig_small}). The right plot shows the yield integrated  $>5.0$ GeV/c in $p_{T}$, for two A+A data sets and two p+p datasets. In both plots the integration is carried out for the data, p+p fit and pQCD  curves from Fig.\,\ref{fig:fig_scaling2}.
}
\label{fig:fig_scaling3}
\end{figure}

Now in order to quantify the direct photon spectra, we first integrate the $p_{T}$ spectra above $p_{T} = 1$\,GeV/c and obtain the left plot of Fig.\,\ref{fig:fig_scaling3}. This plot is another representation of the direct photon scaling, where the integrated yield from the large systems scales with $dN_{ch}/d\eta$ by the same power $\alpha = 1.25$, i.e., $dN_{\gamma}/dy$ grows faster than $dN_{ch}/d\eta$. Also, we show the integrated yield of extrapolations (extrapolated down to $p_{T} = 1$\,GeV/c) of the fit to p+p data and of the three different pQCD calculations scaled by $N_{coll}$. It is quite interesting that the prompt photons (the purple band) and integrated pQCD curves have nearly the same slopes as that of the large systems.

\vskip 0.0truecm

By integrating above $p_{T} = 5$\,GeV/c, we get the right plot of Fig.\,\ref{fig:fig_scaling3}. Here the observed scaling behavior is expected (since $R_{AA} =1$ \cite{Afanasiev:2012}), though we see that the slopes are almost the same as those in the left plot. The integrated pQCD yields scaled by $N_{coll}$ are also shown. The black dashed lines are the power-law fits over the A+A data with a fixed $\alpha=1.25$ slope.

\vspace{0.0\textwidth}
\section{Concluding remarks and summary}
\label{sum}
The PHENIX collaboration has measured low momentum direct photons in Cu+Cu (\cite{Adare:cucu}) and Au+Au at \sqn = 200\,GeV, in Au+ Au at \sqn = 62.4 and \sqn = 39\,GeV as well as in p+p, p+Au and d+Au at \sqn = 200\,GeV. Considering all the available data on small and large systems at various energies, we observe a surprising scaling behavior of direct photons in large systems, namely: at a given center-of-mass energy the low- and high-$p_{T}$ direct photon invariant yields from A+A collisions scale with $N_{coll}$; then $N_{coll}$ is proportional to $dN_{ch}/d\eta$ across different energies; meanwhile, for all energies the low-$p_{T}$ yield scales like $(dN_{ch}/d\eta)^{\alpha}$. PHENIX has also discovered direct photon excess yield (within systematic uncertainties) at low-$p_{T}$ in central p+Au collisions above $N_{coll}$ scaled p+p fit, which may originate from possibly existing QGP droplets in small central systems.  In the low $dN_{ch}/d\eta$ region of the left plot of Fig.\,\ref{fig:fig_scaling3} we see a gradually increasing trend of the integrated yield from small systems, which seems to intersect with a trend from large systems. Both observed trends suggest the existence of a ``thermal transition region or point"  between small and large systems.









\end{document}